\documentclass[12pt,preprint]{aastex}

\newcommand{\EV}[1]{\langle #1 \rangle}
\newcommand{\ecos}{\EV{\cos^{2}(\theta)}}
\newcommand{\e}{\epsilon}
\newcommand{\ecosm}{\EV{M_{i}M_{j}\cos^{2}(\theta)}/\EV{M_{i}M_{j}}}
\newcommand{\ecose}{\EV{\e_{i}\e_{j}\cos^{2}(\theta)}/\EV{\e_{i}\e_{j}}}
\newcommand{\dist}{h^{-1}\,{\rm Mpc}}
\newcommand{\mass}{h^{-1}\,{\rm M_{\sun}}}
\newcommand{\lcdm}{$\Lambda${\rm CDM}}
\newcommand{\mF}{\rm M_{\rm FOF}}

\newcommand{\tc}{\theta_{c}}
\newcommand{\tp}{\theta_{p}}
\newcommand{\ecosc}{\EV{\cos^{2}(\tc)}}
\newcommand{\ecosp}{\EV{\cos^{2}(\tp)}}
\newcommand{\dlin}{n_{fil}}

\shorttitle{\lcdm\ Cluster Alignments $\&$ Ellipticities}
\shortauthors{Hopkins, Bahcall, $\&$ Bode}
\begin{document}

\title{Cluster Alignments and Ellipticities in \lcdm\ Cosmology}
\author{Philip F. Hopkins\altaffilmark{1}, Neta Bahcall\altaffilmark{1}, 
Paul Bode\altaffilmark{1}}
\altaffiltext{1}{Princeton University Observatory, Princeton, NJ 08544, USA}

\begin{abstract}
The ellipticities and alignments
of clusters of galaxies, and their evolution with redshift, 
are examined in the context of
a $\Lambda$-dominated cold dark matter cosmology. 
We use a large-scale, high-resolution N-body simulation
to model the matter distribution in a light cone
containing $\sim 10^{6}$ clusters of mass ${\rm M}>2\times10^{13}\,\mass$ 
out to redshifts of $z=3$.
The best-fit three-dimensional ellipsoid of
the mass distribution is determined for each cluster,
and the results are used to 
analyze cluster ellipticities as a function of mass, radius, and redshift. 
A similar analysis is done
in two dimensions in order to allow direct comparisons with
future observations. Cluster ellipticities are determined within different radii, 
including 0.5, 1.0, and 1.5 $h^{-1}$ comoving Mpc.
We find strong cluster ellipticities: $\EV{\e}\equiv\EV{1-a_{2}/a_{1}}\sim0.3-0.5$. 
The mean ellipticity increases with redshift from $\EV{\e}\sim0.3$
at $z=0$ to $\EV{\e}\sim0.5$ at $z=3$, for both 3D and 2D
ellipticities; the evolution is well-fit by $\EV{\e}=0.33+0.05 z$.
The ellipticities increase with cluster mass and with cluster radius; the main 
cluster body is more elliptical than the cluster cores, 
but the increase of ellipticities with redshift is preserved.
Using the fitted cluster ellipsoids, we determine the alignment of clusters as a function
of their separation. We find strong alignment of clusters for separations $\lesssim100\,\dist$;
the alignment increases with decreasing separation
and with increasing redshift. The 
evolution of clusters
from highly aligned and elongated systems at early times
to lower alignment and elongation at present
reflects the hierarchical and filamentary
nature of structure formation. 
These measures of cluster ellipticity and alignment
will provide a new test of the current cosmological model
when compared with upcoming cluster surveys.
\end{abstract}


\section{Introduction}
In this paper we
examine two aspects of cluster structure that have 
received some, but not extensive,
attention. The first is the distribution of cluster ellipticities and 
the evolution of this distribution with 
mass and redshift. 
The second is
cluster alignment: how aligned are clusters with their neighbors, and how does the
alignment change with cluster mass, separation, and redshift?
Ongoing optical and X-ray surveys, and surveys based on weak
gravitational lensing and the S-Z effect, will provide larger
samples of clusters than currently available; these surveys will enable the measurement
of cluster ellipticities and alignments. When compared with theoretical expectations
as provided below, these measures,
and their evolution with redshift, will provide a test of the cosmological 
model and help our understanding
of cluster formation and evolution.

Observations of galaxy clusters suggest that their shapes, 
as traced by the distribution of galaxies or X-ray emission within the
cluster, are not usually spherical but instead tend to be elongated
\citep{carter80,west89,plionis91,detheije95,basilakos00}.
\citet{melott01} and \citet{plionis02} studied clusters to $z\leq0.13$
and suggest evolution in the gross morphology of clusters in this range,
but \citet{rsmm04} have pointed out that different cluster samples
do not give consistent rates of evolution.
\citet{floor03} investigated simulated clusters at $z<0.2$ and found 
little evolution in this narrow range;
\citet{JS02}, \citet{suwa03}, and \citet{rsmm04}
found evidence of ellipticity evolution in \lcdm\ 
simulations out to $z=0.5-1$.
The evolution of cluster ellipticities to higher
redshifts has not been investigated. 

Conventional models suggest that galaxy clusters 
form through a process of hierarchical clustering, where gas and 
galaxies flow into denser regions along interconnecting large-scale filamentary structures
\citep[e.g.,][and references therein]{shandarin84,gott96,ostriker96}. 
As a result of these inflows, we expect clusters to be aligned with 
their neighbors, especially when both are members of the same filament.
Previous observations of cluster alignments have yielded controversial results.
\citet{binggeli82} found that Abell clusters tend to align with their nearest neighbor
out to separations of $\sim15\,\dist$, and that cluster orientation is related to the
distribution of all surrounding clusters out to separations of $\sim50\,\dist$. 
However, contradictory results have been claimed by others;
for example, \citet{struble85} extended 
Binggeli's method and failed to find significant alignment.
\citet{flin87}, \citet{rhee87}, \citet{west89}, and \citet{plionis94} found optical 
alignment for scales below $30\,\dist$. X-ray emission from clusters 
has also been used
to determine cluster shapes and alignments. 
\citet{ulmer89} studied 46 {\it Einstein}
clusters and did not find alignment; but \citet{chambers00} re-examined their 
results using updated cluster positions and found alignment
among near neighbors. \citet{rhee92}
combined X-ray and optical data, and found that members
of the same supercluster tend to be aligned. 
\citet{west95} found correlation 
among cluster X-ray orientations out to $\sim10\,\dist$, and 
\citet{chambers02} used
{\it Einstein} and {\it ROSAT} data to find alignment
for separations below $30\,\dist$. 

Advances in computational power enable large-volume, high-resolution cosmological
simulations that provide a large sample of clusters for studying
ellipticities and alignments and their evolution with redshift. Previous 
simulations by
\citet{splinter97}, \citet{fgkm02}, and \citet{onuora00}
used $128^{3}, 256^3$, and $512^{3}$ dark-matter particles, respectively, 
to study 
ellipticities and orientations of nearby ($z\sim0$) clusters;
all found significant alignments for nearby clusters.
In this paper
we use a large-scale high-resolution simulation of the current 
best-fit \lcdm\ cosmological model, 
with $2\times10^{9}$ particles, to study 
cluster ellipticities, orientations, and 
alignments in much greater detail than previously possible. 
We examine cluster alignments as a function of 
cluster pair separation and find significant structure in this 
relation for separations $\lesssim100\,\dist$,
and also determine the evolution of cluster ellipticity and alignment 
with redshift, from $z=0$ to $z=3$.
This work seeks to provide a foundation for comparison with future observations
of large cluster samples to high redshifts. 
Such comparison can be used to test the predictions of the current cosmological model as well as 
enhance our understanding of cluster formation and evolution.
In order to provide easy comparison with observations, all
important quantities, such as ellipticity and alignment measures, are obtained using both
the three-dimensional halo dark matter particle distribution and the projected
two-dimensional particle distribution; the latter provides a direct comparison with observations. 
To further facilitate comparison with observations, our analyses of ellipticity and cluster alignment 
are repeated using fixed cluster radii typically
used in observations, rather than the friends-of-friends (or virial) radius.

We describe the simulation and cluster selection in \S\ref{sec:sim}.
The derived cluster ellipticity distribution and its dependence on cluster mass and 
redshift are presented in \S\ref{sec:ecc}.
The alignment of clusters as a function of cluster separation and mass, and the evolution of 
alignment with redshift, are discussed in 
\S\ref{sec:align}. Our conclusions are
summarized in \S\ref{sec:conclusions}.

\section{The Simulation\label{sec:sim}}
We use a high-resolution N-body simulation of
a flat universe with
matter density $\Omega_{m}=0.27$ (baryon density $\Omega_{\rm b}h^2=0.023$ plus cold dark matter), 
cosmological constant $\Omega_{\Lambda}=0.73$, Hubble constant 
${\rm H}_{0}=h\,100\,{\rm km\,s^{-1}\,Mpc^{-1}}=70\,{\rm km\,s^{-1}\,Mpc^{-1}}$,
primordial scalar spectral index $n_s=0.96$, and linear matter power spectrum amplitude
$\sigma_{8}=0.84$. These values are consistent with both the cosmic microwave background 
fluctuation spectrum of WMAP as well as with large-scale galaxy 
and cluster results \citep[e.g.,][]{bahcall99,SpergWMAP,tegmark04}.

The simulation contains $1260^{3}\approx 2$ billion 
particles in a periodic cube. 
The box length is $1500\,\dist$ and the individual particle mass is 
$1.264\times10^{11}\,\mass$. The spline kernel softening
length is $\epsilon=17\,h^{-1}\,{\rm kpc}$; thus, the spatial resolution is comfortably smaller than
the $\sim 100\,h^{-1}\,{\rm kpc}$ characteristic core size of clusters, and the high mass resolution
ensures that two-body relaxation is unimportant in the cluster cores.
The initial conditions were generated using the publicly available codes GRAFIC2 and
LINGERS\footnote{These codes are available at
\url{http://arcturus.mit.edu/grafic/}}
\citep{Bert01, MaBert95}.
The simulation was carried out using the TPM (Tree-Particle-Mesh)
code\footnote{Publically available at
\url{http://astro.princeton.edu/$\sim$bode/TPM/}}
\citep{BO03}, using a PM mesh of size 1260$^3$.
The opening angle in the tree portion of the code was
$\theta=0.577\approx 3^{-1/2}$.  The time step parameter was
set to $\eta=0.3$, and the initial domain decomposition
parameters were $A=2.0$ and $B=12.0$
\citep[for details on these see][]{BO03}.
The computation was carried out using 420 processors on the
Terascale Computing System at the Pittsburgh Supercomputing Center,
and the running time was approximately five days. 
At the end of the simulation $\sim10^6$ trees--- occupying 2\%
of the volume (including all PM cells with an overdensity
above 39) and containing 40\% of the mass--- were being followed
at full resolution.
In the following, only objects containing 160 or more particles,
i.e. with masses above $2\times 10^{13}\,\mass$, will
be considered; this is four times above the limit imposed
by the densest PM-only cell, thus we can be confident that resolution 
will not cause a problem in our analysis. 
This mass threshold corresponds to typical 
groups of galaxies, and thus includes all clusters, from poor to the richest systems.

Particles were saved at each PM time step in order to later
reconstruct the mass distribution along the light cone.
For distances less than $1500\,\dist$ from the simulation 
origin at $z=0$, all particles in spherical
shells defined by each PM time step 
were saved (although only for distances less than
half the box size, or $z<0.265$, is the shell contained within the
simulation volume).  For distances above $1500\,\dist$, a single
octant with all three spatial coordinates greater than zero
is saved. Data were saved for a total of 670 redshift shells 
in the light cone between $z=3$ and $z=0$.
Since the light cone is larger than the simulation box size,
a given cluster can appear more than once, but at different redshifts.
Cluster halos were identified using a standard FOF (friends-of-friends) 
halo finding routine, run
on the light cone output, using a linking length parameter
$b=0.2$ times the mean interparticle separation, corresponding
to an overdensity of 180 times the mean at all redshifts
\citep{LC94,JFWCCEY01}. Different linking lengths ($b$=0.16 and 0.25,
corresponding to fixed overdensities of 360 and 92)
were also tried for comparison, yielding consistent results.
For comparison, the virial overdensity for spherical
collapse ranges from 97 times the mean at $z$=0 to 174 at $z$=3
for the model used in this paper \citep{BryNor98}.

\section{Cluster Ellipticities and their Evolution\label{sec:ecc}}
We determine the structure of halos in the simulation by fitting each identified halo to a three-dimensional
ellipsoid and recording the lengths and orientations of the three axes of the fitted ellipsoid. First, the center of
each halo was taken to be the halo center of mass. 
Once the center of mass of a given halo was identified, the best-fit ellipse was found by taking the matrix of second
moments of particle positions about the center of mass, defined as 
\begin{equation}
I_{ij}=\Sigma x_{i}x_{j}m, 
\end{equation}
where the sum is over all particles in the halo. The dark matter particles are assumed to 
trace the distribution of galaxies within the cluster, as suggested by observations. 
All particles in our simulation have the same mass, so $m$ was set to 1.  
The normalized eigenvectors of $I$ correspond to the unit vectors of
the axes of the best-fit ellipse, and the corresponding eigenvalues ($\lambda_{i}$) give the lengths of these axes
($a_{i}$) by $a_{i}=\sqrt{5\lambda_{i}}$; 
the factor of $\sqrt{5}$ is merely a geometrical 
normalization factor, which will cancel out in our analysis.
The eigenvalues and corresponding eigenvectors were sorted such 
that $\lambda_{1}>\lambda_{2}>\lambda_{3}$. 
We use the most common measure of cluster ellipticity, defined as 
\begin{equation}
\label{eq:ecc}
\epsilon\equiv1-a_{2}/a_{1}=1-\sqrt{\lambda_{2}/\lambda_{1}},
\end{equation}
where $a_{1}$ and $a_{2}$ are the primary and secondary ellipsoid axes lengths, respectively.
We note that other definitions of ellipticity, e.g. 
$\epsilon\equiv1-(a_{2}/a_{1})^{2}$ 
\citep{detheije95,splinter97}
and eccentricity, $e\equiv\sqrt{1-(a_{2}/a_{1})^{2}}$, 
are used occasionally in some studies.
We highlight these definitions in order to 
clarify comparisons but, unless otherwise specified, we use relation (\ref{eq:ecc}) for the definition
of $\e$ in our analysis, 
as it has the most obvious relationship between ellipticity and axis ratio.
Clearly, $e$ and $\epsilon$ 
increase monotonically with one another inside the physical range $0\leq e \leq 1$, and our results
are not qualitatively changed by using either definition.
We used Monte Carlo simulations to test the algorithm's ability to recover halo ellipticities, and
find that the algorithm recovers ellipticities to within $\pm0.01$, 
with accuracy increasing with higher ellipticity.
The results are insensitive to the number of halo particles (with a slight increase in dispersion
for a lower number of particles). 

In order to facilitate comparison with future observations, the process was repeated by viewing 
halos in two-dimensional projection, as in observations. Because halo orientations are, on average, random, 
we use the projection in the x-y axis. 
The eigenvalues and corresponding eigenvectors were again sorted such 
that $\lambda_{2d,1}>\lambda_{2d,2}$, and the projected ellipticity of each halo was taken to be 
$\e_{2d}\equiv1-a_{2d,2}/a_{2d,1}$. Figure~\ref{ecc2d3d} compares the two and three-dimensional 
ellipticities found for the simulated halos.
The two ellipticities are correlated as $\e\sim \e_{2d}$;
the correlation is stronger
with increasing ellipticity.

We find that clusters have moderate to large ellipticities:
$\EV{\e}\gtrsim0.3$. The distribution of 
ellipticities peaks in 
the range $0.3\lesssim\e\lesssim0.5$, corresponding to an 
axis ratio of $\frac{1}{2}\lesssim a_{2}/a_{1}\lesssim\frac{2}{3}$. Thus clusters
are {\em not} spherical, but elongated ellipsoids. 
This holds for clusters of all masses.
Figure~\ref{eccvsm} shows the
ellipticity as a function of cluster mass in six redshift bins of width $\Delta z=0.5$. 
A weak correlation of ellipticity with cluster mass is observed; 
high-mass clusters tend to have high ellipticities, but the scatter is large. 
This is investigated in more detail below.

The evolution of cluster ellipticities with redshift
is illustrated in Figure~\ref{ecchist}.  The results 
for the 3-D and 2-D ellipticities are presented by the solid and dashed histograms, respectively.
The dotted vertical line shows the mean 3-D ellipticity
at the given redshift. The results reveal that the mean ellipticity increases with redshift from 
$\EV{\e}\sim0.3$ at low redshift to $\EV{\e}\sim0.5$ at $z\sim2.5$. It is also 
clear that despite the dispersion in the $\e_{2d}-\e$ relation, the distributions of 3-D
and 2-D ellipticities are almost identical. 
This is a consequence of secondary ellipticities 
$\e_{\rm sec}=1-a_{3}/a_{2}$
having a similar distribution to primary ellipticities in this range 
(well approximated by a Gaussian with $\EV{\e_{\rm sec}}=0.2$ and 
$\sigma_{\e_{\rm sec}}=0.12$); thus the primary ellipticity $\e$ 
is reproduced on average in projection by triaxial clusters with 
the above mean axis ratios. 
Much of the cluster
population has moderate to large ellipticity ($0.2\lesssim\e\lesssim0.6$), with larger ellipticities at 
higher redshift. Although the dispersion of individual ellipticities about the mean value is large, 
\citet{splinter97} and \citet{floor03} found that, when multiple simulations
of the same cosmology are run,
the mean ellipticity at a given redshift has a very low dispersion, $\sigma_{\rm mean}\lesssim0.03$.
The best-fit Gaussian to the observed distribution of $z\approx0$ cluster ellipticities 
($\EV{\e}\sim0.34, \sigma_{\e}\sim0.16$) from \citet{plionis91} 
is shown for comparison in Figure~\ref{ecchist}; it is consistent with the $z\approx0$ \lcdm\ cluster 
ellipticity distribution (with a slight offset due to the larger mass threshold in the observations).

The evolution of ellipticity with redshift is 
illustrated more clearly in Figure~\ref{eccmeanM}. The mean ellipticity 
of all clusters as a function of redshift is plotted for the 3-D 
ellipticities (bold line). The mean 2-D ellipticity is not shown but is identical to the 3-D at all 
redshifts to within $\pm0.05$. The mean ellipticity 
rises consistently from $\EV{\e}\sim 0.33$ at $z=0$ to $\EV{\e}\sim0.5$ 
at $z=3$. A least-squares linear fit yields 
$\EV{\e}=0.3276(\pm0.0003) + 0.0513(\pm0.0003) z$ (rms $=0.0048$). 
For the projected 2-D ellipticities we find 
$\EV{\e_{2d}}=0.3317(\pm0.0003) + 0.0490(\pm0.0003) z$ (rms $=0.0048$). Thus we find 
an approximately linear increase in mean ellipticity from $z=0$ to $z=3$
with $\EV{\e}=0.33+0.05 z$, for
both the 3-D and 2-D ellipticities.   
This slope is consistent with those 
obtained in previous \lcdm\ simulations at $z<0.5$ \citep{rsmm04}
as well as with some observations at $z<0.1$ \citep{melott01},
although, as pointed out by \citet{rsmm04}, different observational 
catalogs disagree on the slope.

The evolution of ellipticity as a function of cluster 
mass is also presented in Figure~\ref{eccmeanM}.
Clusters are divided into three mass bins: low-mass clusters with 
$2\times10^{13}<{\rm M}<5\times10^{13}\,\mass$, medium-mass clusters with 
$5\times10^{13}<{\rm M}<10^{14}\,\mass$, and high-mass clusters with 
${\rm M}>10^{14}\,\mass$ (where cluster mass is 
the total FOF mass). 
We find, as expected from Figure~\ref{eccvsm}, that higher-mass 
clusters are more elliptical at all redshifts. The ellipticity-redshift relation shows the same trend 
for clusters of all masses, with higher mass clusters shifted to larger mean ellipticity at any given 
redshift. We also find that the projected mean ellipticity traces the 
mean three-dimensional ellipticity in all three mass bins. The mean ellipticity-redshift relation found
for all masses is dominated by the low-mass clusters, as expected, especially at high redshift 
where higher-mass clusters become rare.

The rate of change in ellipticity could be affected by numerical
effects in the cores of the simulated clusters,  particularly
when the number of particles in the cluster is small.  To test
if numerical relaxation is occurring, we repeated the
analysis with a higher mass and force resolution simulation,
evolved with the same code.
The cosmological parameters for this simulation 
are fairly similar to the light cone run
($\Omega_{m}=0.3$, $\Omega_{\Lambda}=0.7$, $h$=0.7, $n_s$=1, and
$\sigma_8$=0.95), but the box size is smaller ($L=320\,\dist$)
and hence the resolution is higher: $N=1024^{3}$, making the
particle mass $2.54\times10^{9}\,\mass$;
the spline softening length is $3.2\,h^{-1}\,{\rm kpc}$. 
The FOF ($b$=0.2) finder was run on the entire simulation cube  at
different redshifts to find
clusters with masses greater
than $2\times10^{13}\,\mass$, (i.e., with at least 7960 particles per cluster, 
as compared to 160 in the primary simulation).
We find no change in our results for 
high or medium-mass clusters, but do find that the slope for low-mass clusters
steepens slightly, to $\frac{d\EV{\e}}{dz}\sim0.06-0.07$. This leaves our 
$z\lesssim1$ results unchanged but implies that ellipticities of low-mass 
clusters at high redshifts may be larger than those in our primary simulation by 
$\sim5-8\%$. 

The clusters studied so far include all dark matter particles within the 
friends-of-friends (FOF) radius. 
However, this does not correspond to an easily observed quantity,
so we repeat the analysis
using typical observed radii of
0.5, 1.0, and 1.5 $\dist$ (comoving), denoted by
$R_{0.5}, R_{1.0}$, and $R_{1.5}$ respectively. 
For these analyses, we use the previously determined center of mass (determined from all halo particles) 
as the center of each halo, and then drop all particles further than the given radius of interest.
The analysis then follows the same methods as above,
but using only the particles within the revised radius. 
In order to ensure that the friends-of-friends halo radius  $R_{\rm FOF}$
extends to the cut-off radius of interest, 
the proper mass threshold must be used for each radius.
The initial mass cut, 
${\rm M}\geq 2\times10^{13}\,\mass$, 
is sufficient to ensure that all halos 
have $R_{\rm FOF}>R_{0.5}$. Similarly, a mass cut of 
${\rm M}>4\times10^{13}\,\mass$
ensures $R_{\rm FOF}>R_{1.0}$, and ${\rm M}>10^{14}\,\mass$
ensures $R_{\rm FOF}>R_{1.5}$. 

The results are presented in Figure~\ref{eccmeanRcut}. 
The figure shows the evolution of the mean ellipticity 
with redshift for both the 3-D and 2-D cases, for cluster radii of
$R_{0.5}$, $R_{1.0}$, and $R_{1.5}$. We find that the measured ellipticities 
(both 3-D and 2-D) decrease strongly for smaller 
radii (for the
same $\rm M_{\rm FOF}$ clusters). As the halo centers are the most dense, 
one expects an increasingly spherically symmetric distribution closer to the center,
due to the shorter relaxation time.  Additionally,
extended structures and filaments which would strongly contribute to measured ellipticity are increasingly
reduced with decreasing radius. 
At $R_{0.5}$, most ellipticities have 
decreased by $\sim 0.1-0.2$ from their $R_{1.5}$ or $R_{\rm FOF}$
values. 
A similar correlation between two and three dimensional ellipticities is 
observed for all radii. 
The best linear fits 
to the evolution of $\e$ with redshift 
for each cut-off radius yield a slope of $\frac{d\EV{\e}}{dz}\sim0.05$, 
with a somewhat shallower slope of $\frac{d\EV{\e}}{dz}\sim0.025$ for $R_{0.5}$. 

To test the dependence of these results on the FOF linking length $b$ used in 
the cluster identification, we identified clusters using two additional
different linking lengths and repeated the analysis.
We find a dependence on linking length similar to the
dependence on cluster radius, as expected. 
Using a linking length $b=0.25$ (enclosing an overdensity of 92)
instead of 
$b=0.20$ (enclosing an overdensity of 180) 
yields clusters with larger FOF radii, well beyond
the virial radius, and thus
systematically increases the mean cluster ellipticities by $\sim0.1$. A linking length of
$b=0.16$ has, as expected, an opposite effect; it systematically decreases 
mean ellipticities by $\sim0.025$, similar to 
reducing the radius to $R_{1.0}$.
In both cases the dependence on 
redshift is preserved.
We note that the fixed radii used above do not
correspond to the virial scales of clusters,
but would be less difficult to implement observationally.

\section{Cluster Alignment and its Evolution\label{sec:align}}
Using the parameters measured in \S~\ref{sec:ecc} for each halo, we
identify pairs of halos and record 
the separation and alignment measure for each pair. 
We examine two common measures of alignment: first, 
the angle between the two major axes of a pair of clusters (the ``correlation'' angle), 
and second, the angle between a given cluster major 
axis and the line connecting the cluster to its paired cluster
(the ``pointing'' angle).
We use as our ``correlation'' alignment measure the 
squared dot product of the two halo major axes (given as unit vectors by the
eigenvectors found in \S~\ref{sec:ecc}), yielding $\cos^{2}(\tc)$ for each pair, where
$\tc$ is the angle between the two major axes. 
We consider $\cos^{2}(\tc)$ instead of
$\cos(\tc)$ because $\pm\cos(\tc)$ are physically identical, since the unit vectors defining
the halo axes do not have a preferred ``positive'' direction. We similarly
use $\cos^{2}(\tp)$ as our ``pointing'' angle measure, but in this case 
$\tp$ is the angle between a cluster major axis and the line connecting its 
center of mass to the center of mass of its paired cluster. We find, as \citet{splinter97},
that both measures give similar results, with a stronger signal for the pointing 
angle measure. We also find that the pointing angle signal extends to larger scales.

Using the correlation angle measure, we find that pairs of clusters are strongly aligned
for separations $\lesssim30\,\dist$. On larger scales, alignment 
between pairs of clusters is random and uniformly distributed.
The alignment function -- the mean cluster alignment 
as a function of cluster pair separation $r$ -- 
is presented 
for different redshift bins
in Figure~\ref{cosvr3d} (for 3-D) 
and Figure~\ref{cosvr2d} (for 2-D).
Assuming a random
uniform distribution in $\theta$ ($P(\theta)=\frac{d\theta}{2\pi}$), 
one expects $\ecos=\frac{1}{3}$
for the 3-D case and $\ecos=\frac{1}{2}$ for the 2-D case
(for both correlation and pointing angle measures); obtaining a
$\ecos$ greater than these values indicates alignment. 
The data yield $\ecosc=\ecos_{\rm ran}$
(where $\ecos_{\rm ran}=\frac{1}{3},\frac{1}{2}$ for 
the 3-D and 2-D cases, respectively) 
for large separations ($r>30\,\dist$).  At smaller separations 
the data show clear alignment of clusters,
with a pronounced rise in $\ecosc$ with decreasing $r$, to
a maximum at $r\approx3\,\dist$. 
The fall in $\ecosc$ below $3\,\dist$ is due to the close proximity of the cluster pairs; clusters at such small
separations begin to overlap, and measures of ellipticity and alignment become less meaningful.
The alignment increases significantly 
with increasing redshift. For example, $\ecosc$ at $r\approx3\,\dist$
rises from $\sim0.40$ at $0<z<0.5$ to $\sim0.46$ at $2.0<z<2.5$. The signal-to-noise ratio decreases at high
redshift, as the number of massive clusters decreases. At the highest redshifts, 
$2.5<z<3.0$, the rise in $\ecosc$ with decreasing
$r$ (to $\ecosc\approx0.48$ at $r\approx3\,\dist$) is considerably steeper than for $z<2.5$.

Similar trends are observed for the 2-D case in Figure~\ref{cosvr2d}. We see 
$\ecosc=\ecos_{\rm ran}=\frac{1}{2}$ for $r\gtrsim20\,\dist$, 
with a similar rise in $\ecosc$ with decreasing $r$.
Alignment increases with increasing redshift,
from $\ecosc_{\rm max}\approx0.54$ for $0<z<0.5$ to $\ecos_{\rm max}\approx0.57$ for $2.0<z<2.5$.
At high redshift, uncertainties are large, and 
for $2.5<z<3.0$ the alignment function becomes more erratic. In this redshift range, 
$\ecosc_{\rm max}\approx0.65$ at $r\approx2\,\dist$. 

Figures~\ref{cospointing3d} and~\ref{cospointing2d} are similar to Figures~\ref{cosvr3d}
and~\ref{cosvr2d}, but use the pointing angle
alignment measure described previously. This measure is the 
angle between a given cluster major 
axis and the line connecting the cluster center of mass to its paired cluster center of mass at the
given separation. We again use $\ecosp$ as our alignment measure, 
employing this alternative definition of
$\tp$. We find that the pointing angle alignment
function is similar to the correlation angle alignment function, with 
larger amplitude and a positive alignment 
signal extending to larger scales of $\sim100\,\dist$.
Previous observations by \citet{west89} and
\citet{plionis94}, which used the pointing angle measure, are shown for comparison in Figure~\ref{cospointing2d}. 
The results are consistent within 1$\sigma$ with the \lcdm\ predicted 2-D alignment at $z=0$.

As in \S~\ref{sec:ecc}, we compare our results 
with those obtained using the more 
readily observed fixed co-moving radii
of 0.5, 1.0, and 1.5 $\dist$. We find that 
the alignment measures are nearly independent (within 
the 1$\sigma$ error bars) of the 
cluster radii used for determining cluster orientations.
A similar near independence holds when varying
the halo linking
lengths; clusters identified using FOF linking lengths of $b=0.25$ and $b=0.16$ 
yield essentially the same results as those using $b=0.20$ 
(with a slight $\sim0.01$ increase in $\ecos$ at small $r$ for $b=0.25$).

We investigate the sensitivity of the alignment function 
to the masses and ellipticities of the halos 
by examining a 
mass and ellipticity-weighted average of $\cos^{2}(\theta)$, 
taken as $\ecosm$ and $\ecose$, where
$M_{i}$ and $M_{j}$ represent the masses of the two 
halos in a given pair (likewise for $\e_{i}$ and $\e_{j}$).
We find that the ellipticity-weighted average closely traces the 
standard expectation value of $\cos^{2}(\theta)$ at all redshifts 
for both the 3-D and
2-D cases, and thus do not further explore a possible ellipticity dependence of this relationship.
We note that \citet{onuora00} also found that highly elongated clusters show no greater tendency 
to alignment than more spherical clusters.

We do find some deviations between $\ecosm$ and $\ecos$, and thus
explore the possible mass dependence of this relation
by dividing clusters into three 
mass bins using their FOF masses: low-mass clusters with 
$2\times10^{13}<{\rm M}<5\times10^{13}\,\mass$, medium-mass clusters with 
$5\times10^{13}<{\rm M}<10^{14}\,\mass$, and high-mass clusters with 
${\rm M}>10^{14}\,\mass$. Because alignment is determined using {\em pairs} of clusters, 
each cluster is labeled
with a letter corresponding to its mass: H (high-mass cluster,
as defined above), M (medium-mass), or L (low-mass). Pairs are then identified as 
either H-H, H-M, H-L, M-M, M-L, or L-L, with the letters corresponding to the two elements of the
cluster pair. We repeat our alignment analysis as a function of redshift for 
each of these six cluster subsamples, using both the correlation angle 
and pointing angle measures of alignment. 

Figure~\ref{cosvsrM} presents the results of this alignment analysis as a function of 
pair separation and redshift for
the three equal-mass pair subsamples, using the pointing angle measure. 
The results show that the cluster alignment function increases and
shifts to larger scales as the mass of the clusters increases from L-L to M-M to H-H. 
The scale at peak alignment of the L-L, M-M, and H-H pairs increases from 
$r\sim2\,\dist$ to $\sim3\,\dist$ to $\sim6\,\dist$, respectively. This is 
unsurprising, as it suggests that more massive cluster pairs show stronger alignments
to larger scales. 
Mixed mass pairs, H-M, M-L, and H-L, follow intermediate
relationships, with smaller peak alignments. 
It appears that clusters of comparable masses yield the strongest alignment; the
alignment strength and scale increases with increasing cluster mass.
Results for the 2-D cluster alignments are presented in Figure~\ref{cosvsrM2d}.
The correlation angle results are qualitatively similar, but with 
weaker amplitude and no significant signal for separations above $\sim50\,\dist$.

The cluster alignment increases for all masses with increasing redshift;
high redshift clusters are more strongly aligned than low redshift clusters. The alignment
increases more rapidly at high redshift among high-mass 
clusters than among low-mass clusters. For $1.0<z<1.5$, 
we find a maximum value of 
$\ecosp\sim0.56$ for L-L pairs, and a maximum value of 
$\ecosp\sim0.65$ for H-H pairs. At $1.5<z<2.0$, 
the maximum alignment of L-L pairs rises to 
$\ecosp\sim0.58$, and the maximum 
alignment of M-M and H-M pairs rises to $\ecosp\sim0.62$ 
(there are too few H-H pairs for statistical analysis). 
At $2.0<z<2.5$, the L-L pairs have a maximum of 
$\ecosp\sim0.6$ and the M-M pairs have a maximum of 
$\ecosp\sim0.7$. The corresponding values for the 
correlation angle alignment function are $\sim25\%$ lower.

These results raise the possibility that
clusters of all masses form with similar large 
alignments. We divide
clusters into six mass bins, and for each mass bin identify the approximate range of 
redshifts in which these clusters formed. We take this range to be the 
range of redshifts over which $\frac{d\log{n}}{dz}\sim10$, the approximate
maximum value of $\frac{d\log{n}}{dz}$ for each mass bin. Here, $n$ is the 
number density of clusters of the given mass in each redshift shell. 
Figure~\ref{compareformation} shows the resulting 
correlation angle alignment function for 
each mass bin. The range of masses and redshifts for each bin are shown in the figure.
We find that clusters of all masses show strong alignment 
shortly after formation, with an alignment function that is nearly independent of mass. 

If cluster alignment is a consequence of cluster formation along filamentary 
superstructures, then the alignment of two clusters is an 
indication of connecting structures.
For each pair of clusters 
with separation $r<50\,\dist$
we compute $N_{ij}$, the number of clusters
inside a cylinder of radius $1.5\,\dist$ along the line
connecting their two centers, and
also the volume $V_{ij}$ inside this cylinder.
The number density of clusters in these ``filaments'', 
$\dlin\equiv {\Sigma N_{ij}}/{\Sigma V_{ij}}$,
can be compared 
to the mean number density of clusters, $\bar{n}$, in any
redshift range. 
The resulting number overdensity in the filaments, $\dlin/\bar{n}$, 
is plotted as a function of 
the mean pointing angle $\cos^{2}(\tp)$ 
(the average of the two pointing angles of the pair) in Figure~\ref{overdense},
for different redshifts. We find that overdensity in the connecting filament 
rises by a factor $\sim5$ at all redshifts as the alignment increases from
$\ecosp=0$ to $\ecosp=1$, indicating that aligned cluster pairs at separations $r<50\,\dist$ 
are more likely to be connected by filamentary superstructures than unaligned 
pairs at the same separation.
The high overdensity 
$\EV{\dlin/\bar{n}}$ at any angle indicates that any two clusters
within $50\,\dist$ are likely to be part of some supercluster. The overall amplitude of
the overdensity is closely related to the 
cluster correlation function, and rises with increasing redshift as expected 
\citep[e.g.,][]{bahcall03}. Similar results are also seen using the cluster mass 
overdensity, or using the fraction of pairs that have another cluster along the connecting
filament.

\section{Discussion\label{sec:conclusions}}
We use a large-scale cosmological simulation of the current best-fit \lcdm\ cosmology to investigate the 
ellipticities and alignments of $\sim10^{6}$ clusters of galaxies. We study the 
dependence of cluster ellipticity and alignment on cluster mass, radius, and redshift.
We find that clusters are elliptical, with mean ellipticities $\EV{\e}\gtrsim0.3$.
The derived distribution of ellipticities at $z\sim0$ is consistent with current observations
\citep{west89,plionis91,detheije95,basilakos00}. 

Our large sample of clusters enables a detailed study of the distribution of cluster ellipticities as a 
function of various cluster properties: mass, radius, and redshift.
We find that the mean cluster ellipticity increases with cluster mass, i.e., massive clusters 
are more elliptical than less massive clusters. We also find that cluster ellipticity 
decreases with radius; cluster cores are less elliptical than cluster outskirts. Both of these 
trends can be understood in terms of infall and mergers; these affect massive clusters and the
outskirts of clusters more significantly than less massive clusters and cluster cores.

We find that the mean cluster ellipticity increases monotonically with increasing redshift 
from $z=0$ to $z\approx3$, 
consistent with previous observation and simulation
at low redshifts \citep{melott01,plionis02,JS02,suwa03,rsmm04}.
This is likely due to hierarchical cluster formation and the high rate of infall and merger at 
early times.
The evolution of cluster ellipticity with redshift does not depend on cluster mass or radius; 
we find $\frac{d\EV{\e}}{dz}\approx0.05$ for 
radii $r\geq1\,\dist$. Below a radius of $1\,\dist$, the 
dependence on redshift weakens as clusters become more isotropic 
closer to the cluster center. Clusters are most elliptical at early times and become 
less elliptical at present; this can be understood as a relaxation process in a low-density 
cosmological background, with a reduced merger rate at later times \citep[e.g.,][]{floor03}. 
Our results suggest that once clusters form, a period of 
relaxation exists without significant new infall or merger, allowing clusters to become more spherical. 

We find that clusters are aligned with one another to separations of at least
$\sim30\,\dist$, 
using both the pointing angle and the correlation angle methods for measuring alignment. 
The pointing angle measure shows a stronger signal at all separations, and a 
$1\sigma-2\sigma$ alignment detection to $\sim100\,\dist$ (compared to $\sim30\,\dist$
for the correlation angle measure). 
The pointing angle emphasizes the alignment of clusters with 
filamentary structures along which nearby clusters would form, and the correlation
angle, while similar, emphasizes the direct alignment of clusters with one another; 
the greater strength of the pointing angle signal suggests that alignment is related to
filamentary structures. If individual cluster alignments are independently 
perturbed, as is likely, these perturbations will dilute the correlation angle signal 
more strongly than the pointing angle signal.

The alignment increases with decreasing cluster
separation, revealing a smooth alignment function.
The alignment function shows a steady increase in alignment strength 
from large separations of $\sim100\,\dist$ with the pointing angle measure
(or $\sim30\,\dist$, with the correlation angle measure)
to a separation of $r\sim3\,\dist$, where maximum alignment is seen. 
At smaller separations, the alignment measure decreases 
as clusters begin to overlap and alignments and ellipticities cannot be 
reliably measured. Distortions due to mergers and 
difficulty distinguishing elements of two clusters at such small separations
render this region less meaningful. 
The alignment is independent of the cluster radius used in the
analysis.
We find that cluster alignment is 
correlated with the existence of filamentary 
structure along the line connecting a pair of clusters, with 
the cluster number overdensity increasing by a factor of $\sim5$ as
the alignment increases from no alignment to maximum alignment. The highly aligned 
clusters are therefore more typically members of filamentary superclusters. 

We find that cluster alignment evolves with redshift. 
The alignment increases for all cluster separations with
increasing redshift from $z=0$ to $z\approx3$. 
Objects are most aligned at early times; we find 
that $\ecos_{\rm max}$ decreases from $\sim0.55$ at $z\sim3$ to $\sim0.4$ 
at $z\sim0$. Our results suggest that alignment is a 
consequence of initial structure formation, and decreases rapidly over a timespan
$\Delta z\sim0.5-1$ after which alignments continue to become 
more random, but at a slower rate. This supports the picture that 
alignment stems from early large-scale structure, with cluster formation by hierarchical
clustering in which material falls into the cluster along large-scale filamentary structures
\citep[e.g.,][]{onuora00}. This evolution of cluster alignment with redshift 
provides a new test of the cosmological model, even if present ($z=0$) alignments do not
significantly differ between models. 

Cluster alignment increases with cluster mass. The alignment extends to larger scales 
as the paired masses increase from low to medium to high-mass 
(L-L to M-M to H-H pairs). The alignment weakens somewhat for mixed-mass 
pairs (at all redshifts). 
At high redshifts $z\gtrsim1$, alignments of high-mass clusters increase faster than 
alignments of low-mass clusters. We find that clusters of all masses 
form with similar strong alignments, 
and that alignment rapidly declines to a steady value over a 
timespan $\Delta z\sim0.5-1$. Most low-mass objects are formed early and thus decline to 
a steady value ($\ecos\sim0.42-0.44$) before high-mass objects can form and do so. 
This rapid evolution 
may be the result of a period of violent relaxation during which dissipation 
and gravitational instability increase isotropy and randomize cluster alignments. 
The strong evolution of the alignment signal seen shortly after 
cluster formation suggests that
observations of cluster alignments at the characteristic redshifts of their formation, for 
the relevant cluster masses, may discriminate more strongly between cosmological models
 than observations at $z=0$.

Cluster alignments and ellipticities both show 
dependence on cluster mass and redshift. The distribution of ellipticities and 
alignments can inform our models of large-scale structure 
formation, and can provide a new and independent test of the cosmological model. 
Our results provide
detailed predictions of the current cosmology 
for the ellipticity distribution and alignment function of clusters 
as a function of cluster mass, radius, and redshift; these predictions can be used
for comparison with upcoming 
observations of large cluster samples to high redshifts. 

\acknowledgments

We thank J.P. Ostriker, David Spergel, and Michael Strauss
for helpful discussions.
This research was supported in part by NSF grant 0407305.
Computations were performed on the National Science Foundation
Terascale Computing System at the Pittsburgh Supercomputing Center;
computational facilities at Princeton were provided by 
NSF grant AST-0216105.  PB acknowledges support from NCSA under 
NSF Cooperative Agreement ASC97-40300, PACI Subaward 766.

\begin{center}
\begin{figure}
    \plotone{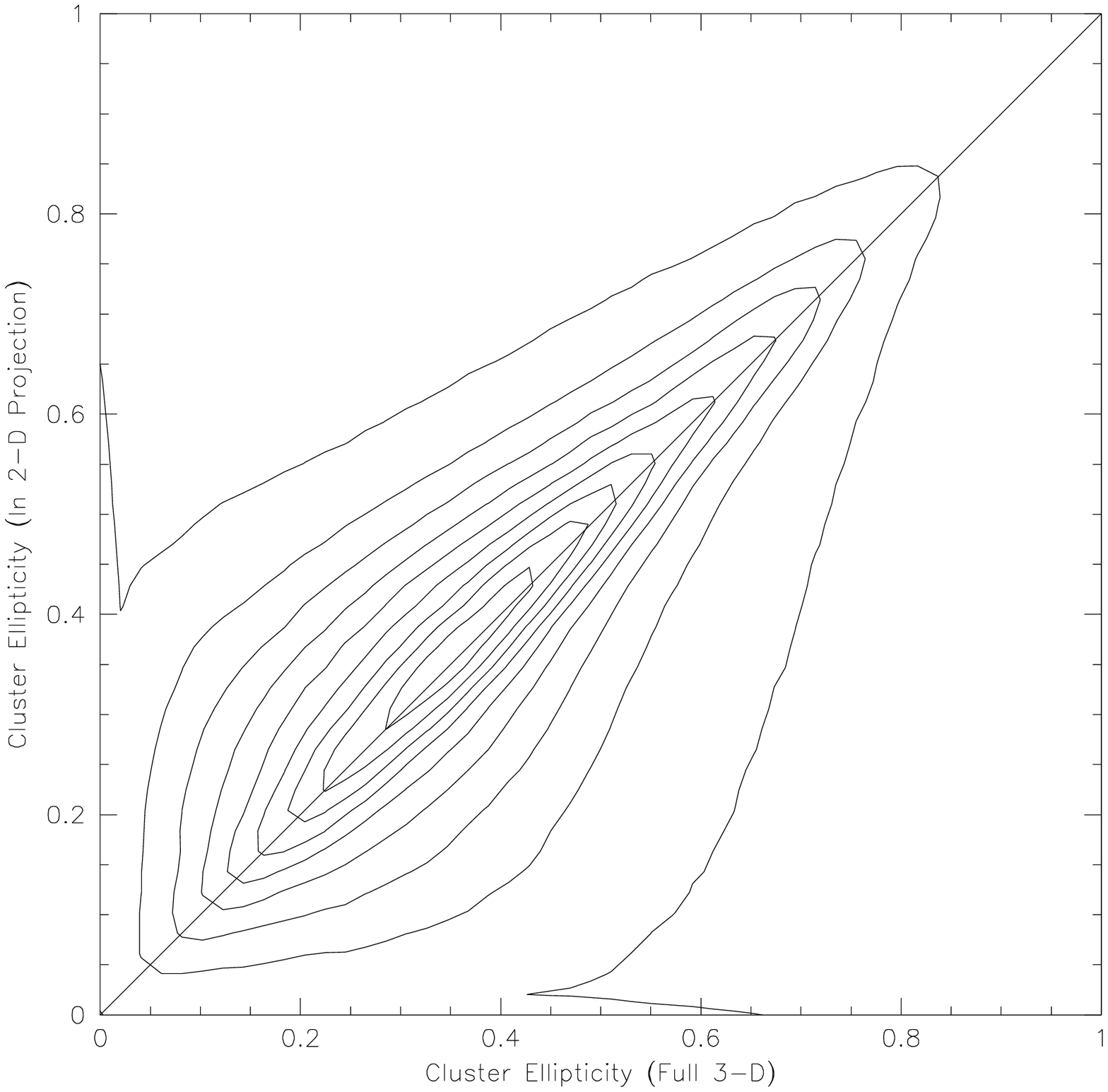}
    \caption{Density contours at $11\%$ intervals show the three dimensional 
    vs.\ the two dimensional (projected) ellipticity (obtained from the best-fit
    ellipse to the 2-D projection of particle positions)
    for all clusters with mass $\mF\geq 2\times10^{13}\,\mass$ (at all redshifts).
    The correlation follows $\e\sim \e_{2d}$.
    \label{ecc2d3d}}
\end{figure}
\begin{figure}
    \plotone{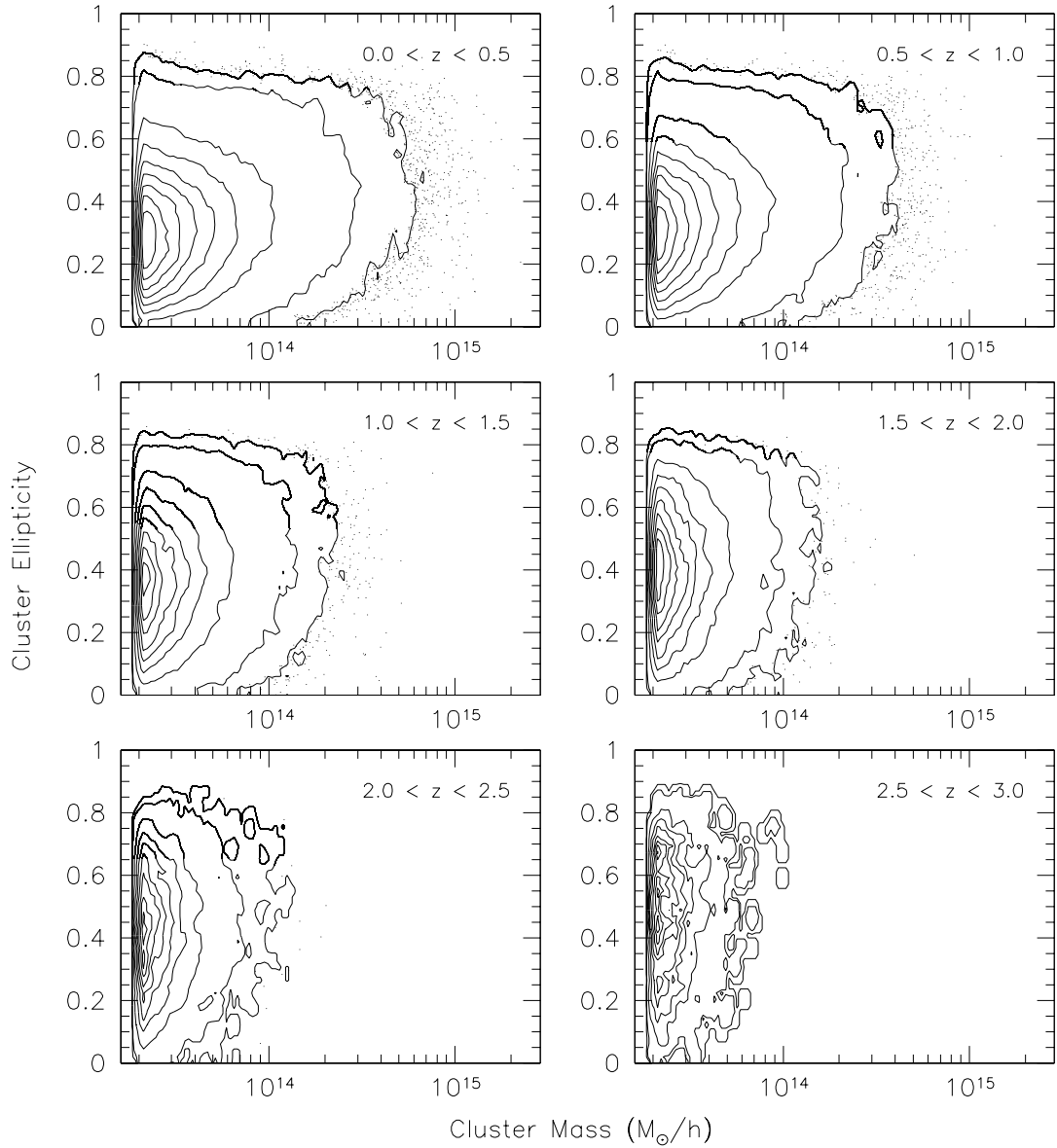}
    \caption{Cluster ellipticity as a function of cluster mass in six redshift bins,
    with density contours at $11\%$ intervals. 
    \label{eccvsm}}
\end{figure}
\begin{figure}
    \plotone{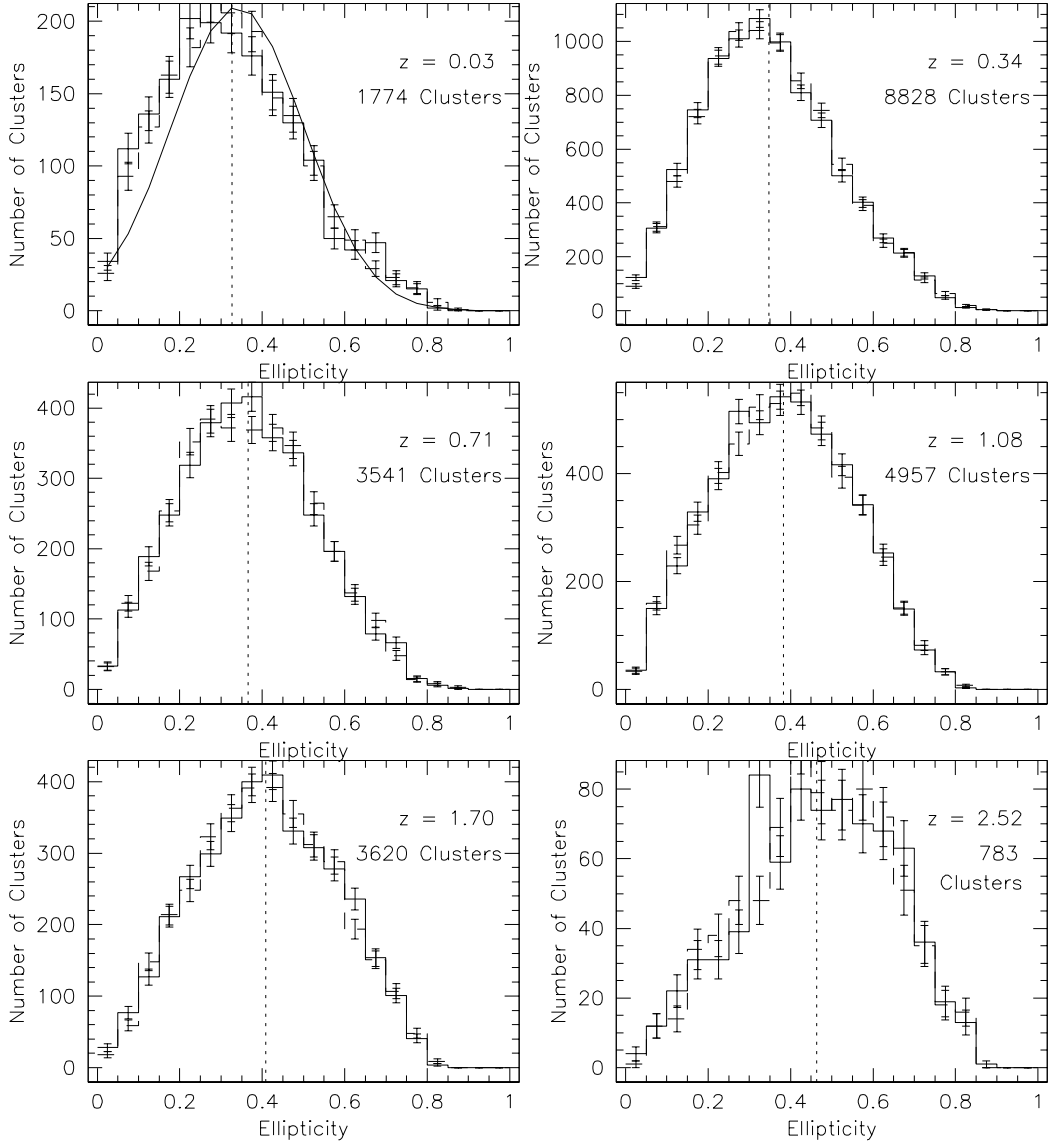}
    \caption{Histograms of the ellipticity distribution at the listed 
    redshifts. The solid histogram is for the three-dimensional ellipticities ($\e$), and
    the dashed histogram is for the projected two-dimensional ellipticities ($\e_{2d}$). The 
    dotted line shows the mean 3-D ellipticity within the redshift step. Error bars shown
    assume Poisson errors ($\sigma=1/\sqrt{n}$).
    The curve in the upper left shows the best-fit Gaussian to the observed distribution of $z\approx0$ Abell 
    cluster ellipticities \citep{plionis91}; the observed clusters are more
    massive than the threshold used here, causing the slight shift to higher ellipticities. 
    \label{ecchist}}
\end{figure}
\begin{figure}
    \plotone{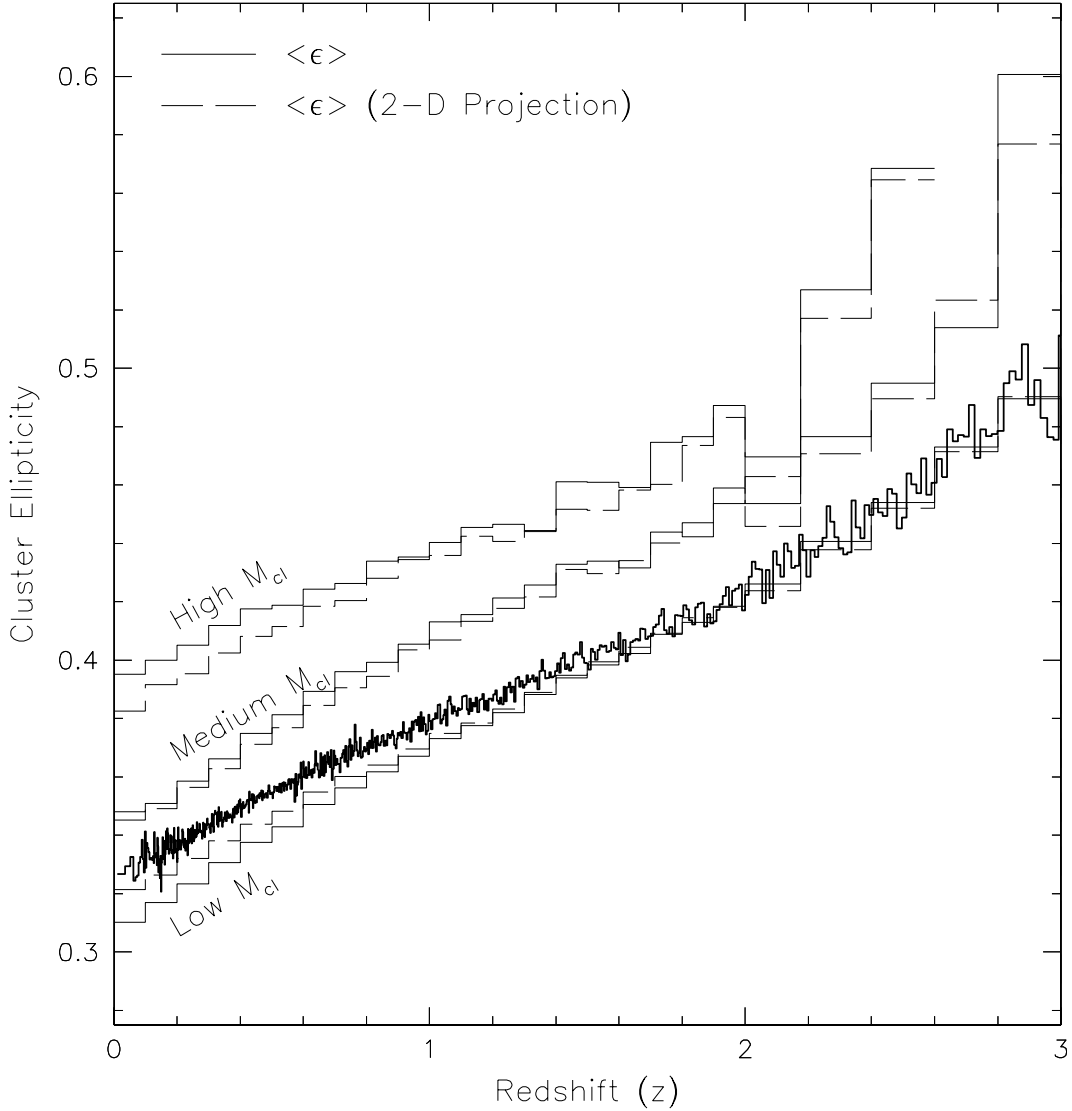}
    \caption{Evolution of mean cluster ellipticity as a function of redshift 
    for different mass clusters (M$_{cl}$, within the FOF radius). 
    The bold line is the 3-D mean 
    ellipticity $\EV{\e}$ of clusters of all masses (${\rm M}\geq2\times10^{13}$)
    (the difference between the 3-D and 2-D mean ellipticities is negligible). 
    The lower lines show the relationship for low-mass 
    ($2\times10^{13}\leq {\rm M}<5\times10^{13}\,\mass$) 
    clusters, middle lines for medium-mass 
    ($5\times10^{13}<{\rm M}<10^{14}\,\mass$) clusters, and upper lines for 
    high-mass (${\rm M}>10^{14}\,\mass$) clusters. For each, the solid line is the mean three-dimensional
    ellipticity $\EV{\e}$, and the dashed line is the projected two-dimensional ellipticity $\EV{\e_{2d}}$.
    \label{eccmeanM}}
\end{figure}
\begin{figure}
    \plotone{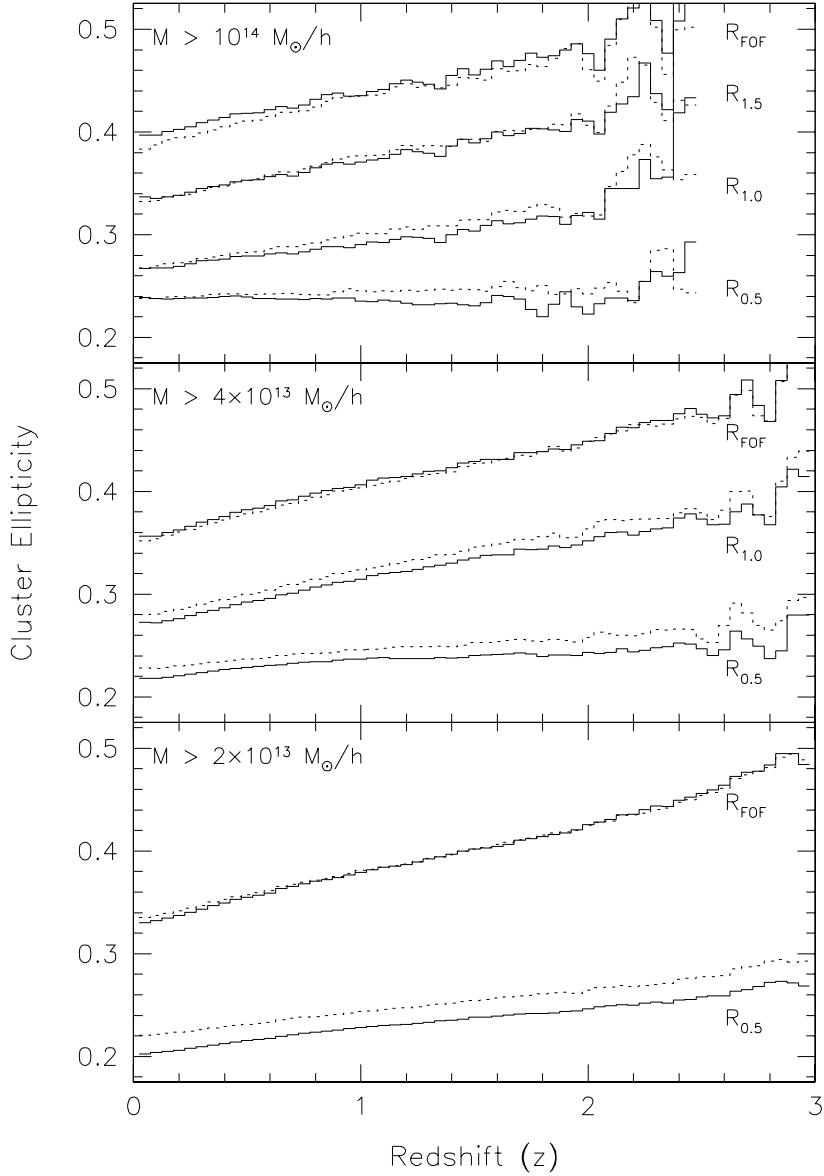}
    \caption{Mean cluster ellipticity as a function of redshift for different cluster radii
    ($R_{\rm FOF}, R=1.5, 1.0, {\rm and}\ 0.5\,\dist$). Solid and dotted lines
    represent the 3-D and 2-D ellipticities, respectively.
    The top panel shows the results for all clusters massive enough to ensure $R_{\rm FOF}>1.5\,\dist$, 
    the middle panel for $R_{\rm FOF}>1.0\,\dist$, and the lower panel for $R_{\rm FOF}>0.5\,\dist$.
(The slight increase of 2-D over 3-D ellipticity for $R_{0.5}$ is due 
to the secondary ellipticity becoming more dominant on small scales,
where the primary ellipticity is small;  the 2-D distribution also
includes projected material outside the 3-D radius, 
which can further increase ellipticity.)
    \label{eccmeanRcut}}
\end{figure}
\begin{figure}
    \plotone{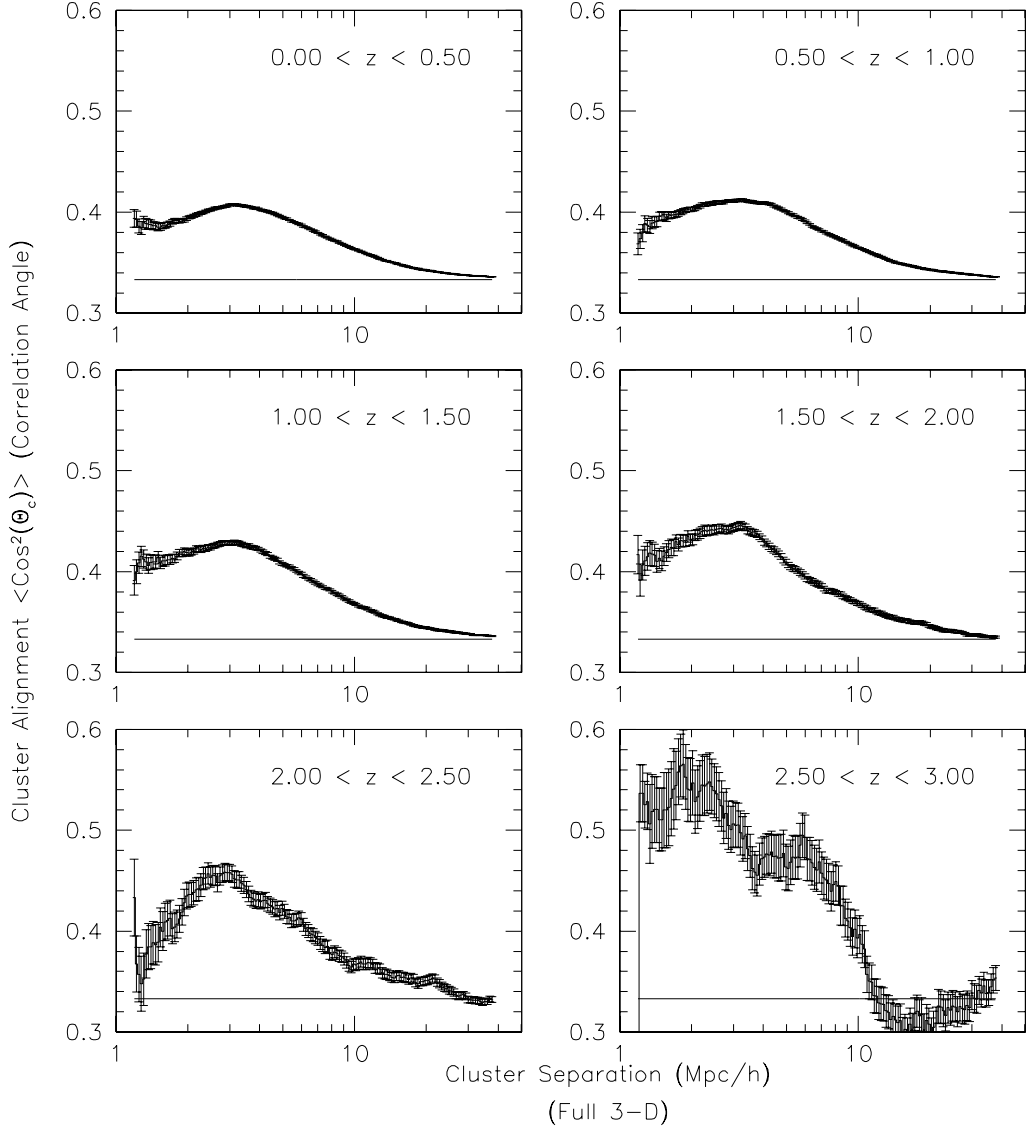}
    \caption{The 3-D cluster alignment function (i.e.,
    alignment as a function of halo separation) using the correlation angle method
    is shown for six redshift bins. 
    Error bars represent $1\sigma$ uncertainties. The alignment measure, $\ecosc$, is 
    determined within bins of 
    $\pm0.1\log_{10}(r)$, where $r$ is the cluster pair separation in $\dist$. 
    The horizontal line indicates the expected
    value for random orientations, $\ecos_{\rm ran}=\frac{1}{3}$.
    \label{cosvr3d}}
\end{figure}
\begin{figure}
    \plotone{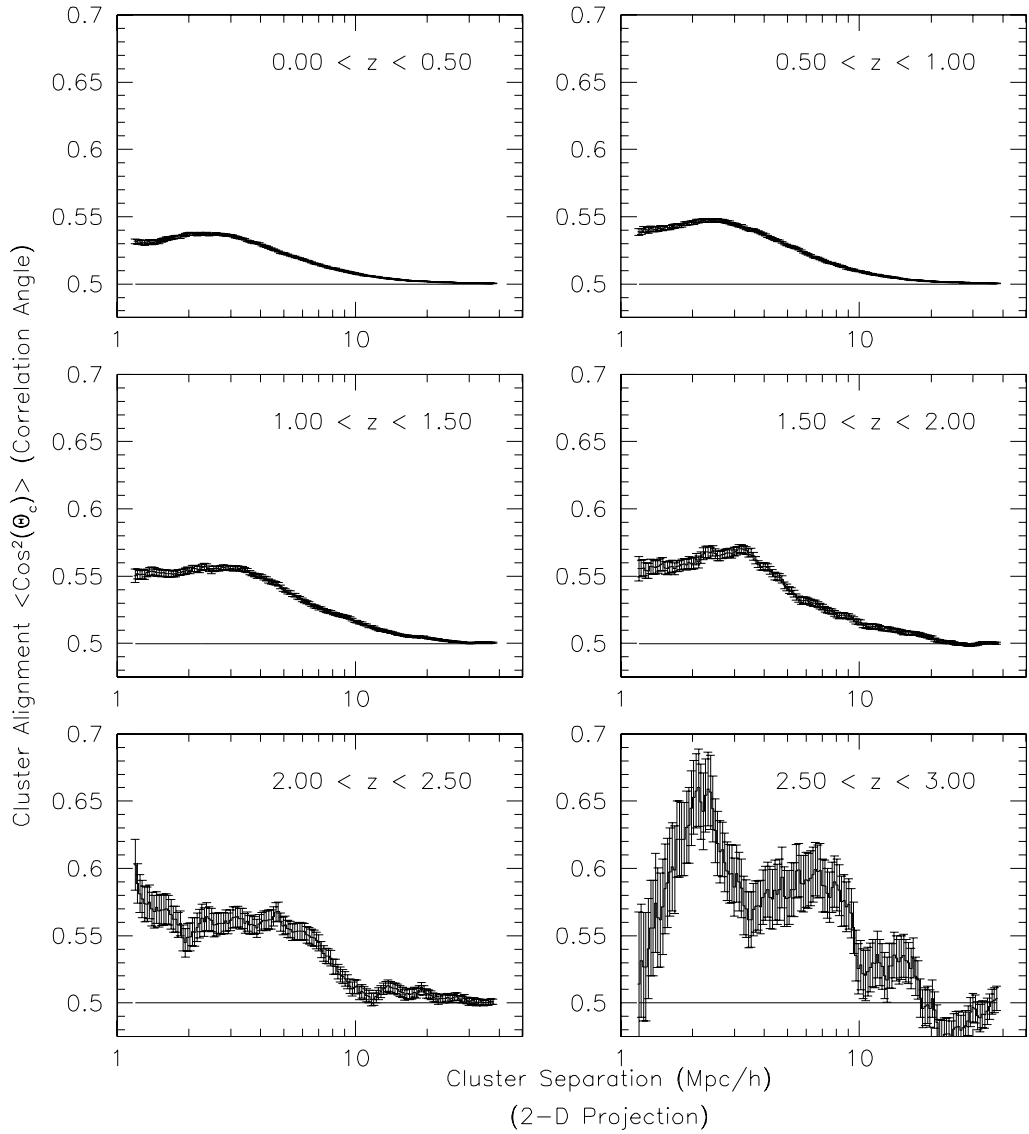}
    \caption{Same as Figure~\ref{cosvr3d}, but for the 2-D correlation 
    angle alignment function. 
    \label{cosvr2d}}
\end{figure}
\begin{figure}
    \plotone{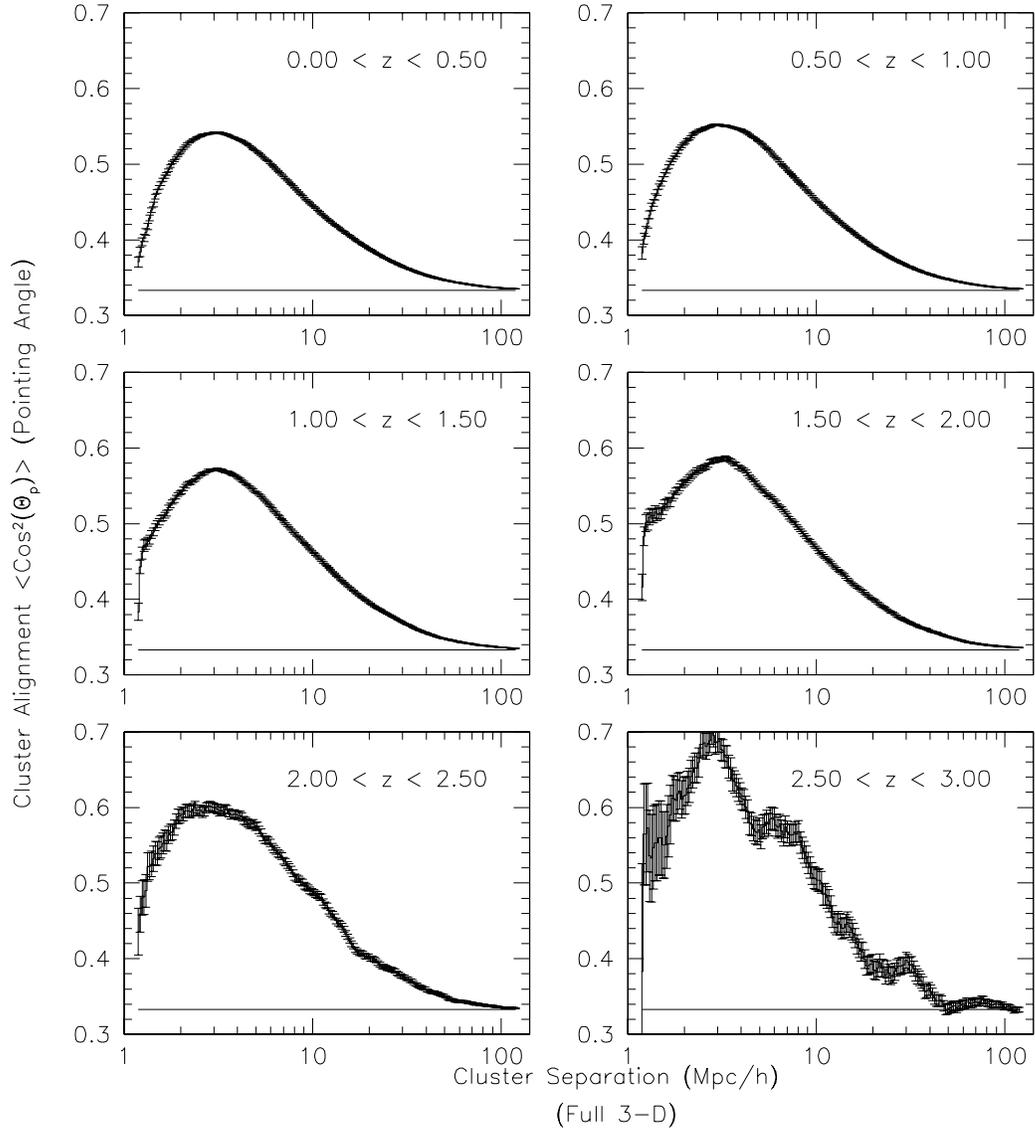}
    \caption{Same as Figure~\ref{cosvr3d}, but for the 
    3-D alignment function using the pointing angle method. 
    \label{cospointing3d}}
\end{figure}
\begin{figure}
    \plotone{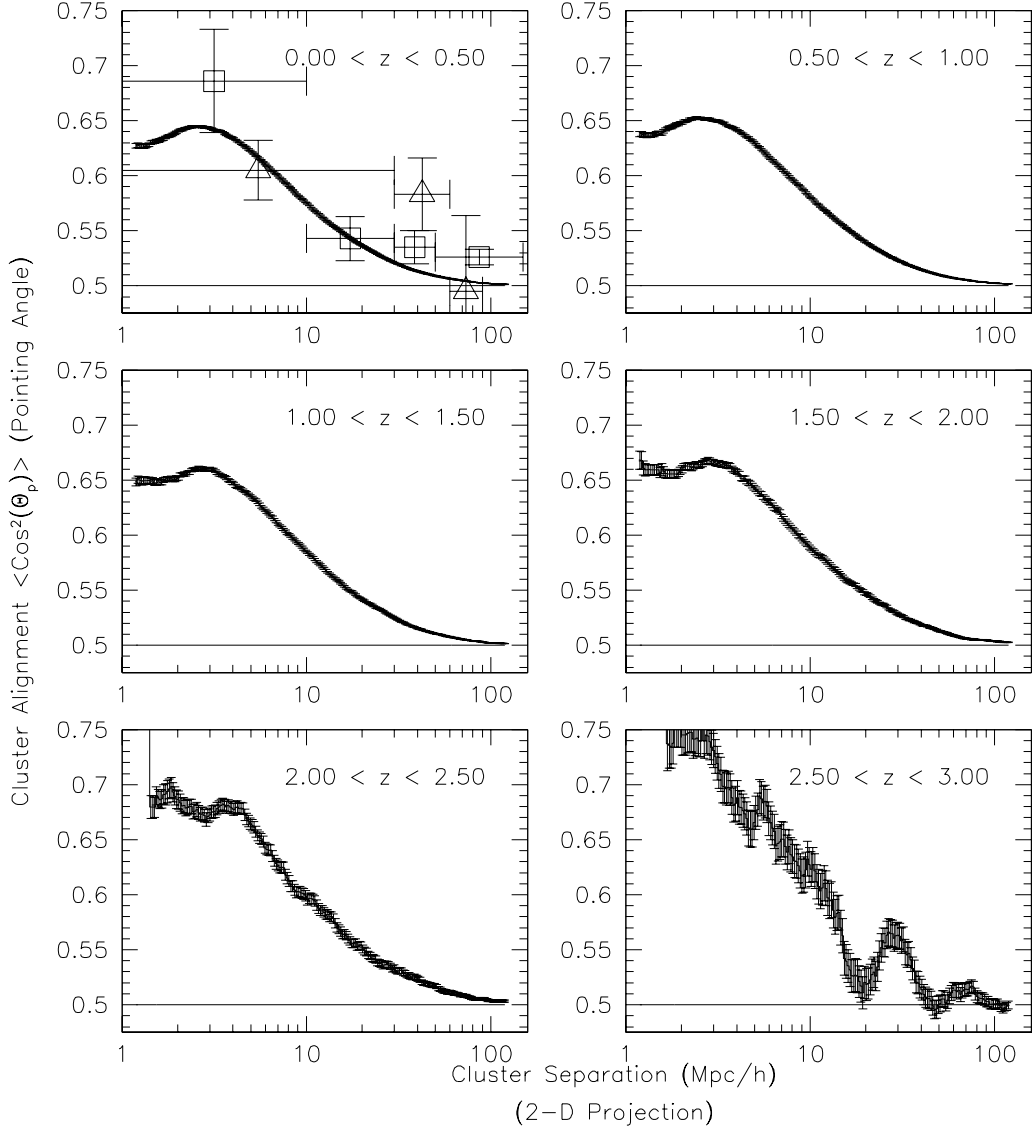}
    \caption{Same as Figure~\ref{cosvr3d}, but for the 2-D 
    alignment function using the pointing angle method. 
    The horizontal line indicates the expected
    value for random orientations, $\ecos_{\rm ran}=\frac{1}{2}$.
    The points show observed alignments (converted 
    to $\ecosp$) from \citet{west89} (triangles) and \citet{plionis94} (squares), 
    for Abell and Schectman clusters at $z\approx0$.
    \label{cospointing2d}}
\end{figure}
\begin{figure}
    \plotone{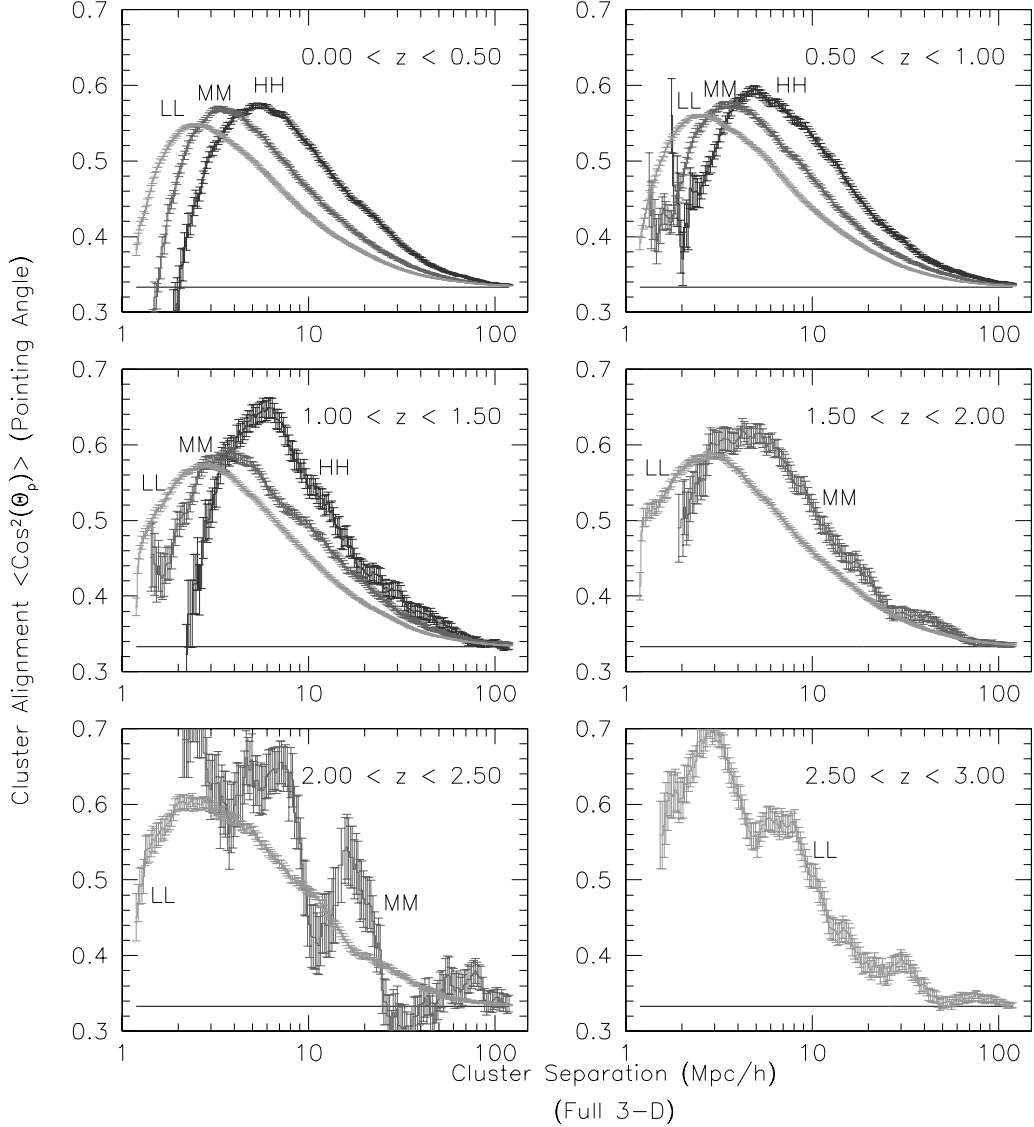}
    \caption{The 3-D cluster alignment function (pointing angle method) 
    in six redshift bins
    (as in Figure~\ref{cospointing3d}), 
    for different mass cluster pairs (see \S~\ref{sec:align}).
    From darkest to lightest, the curves represent H-H pairs, M-M pairs, 
    and L-L pairs, as labeled. At high redshift, the number of high-mass
    cluster pairs is too small for reliable analysis. 
    Mixed-mass pairs are not shown; they have 
    similar, but somewhat weaker, alignment functions.
    \label{cosvsrM}}
\end{figure}
\begin{figure}
    \plotone{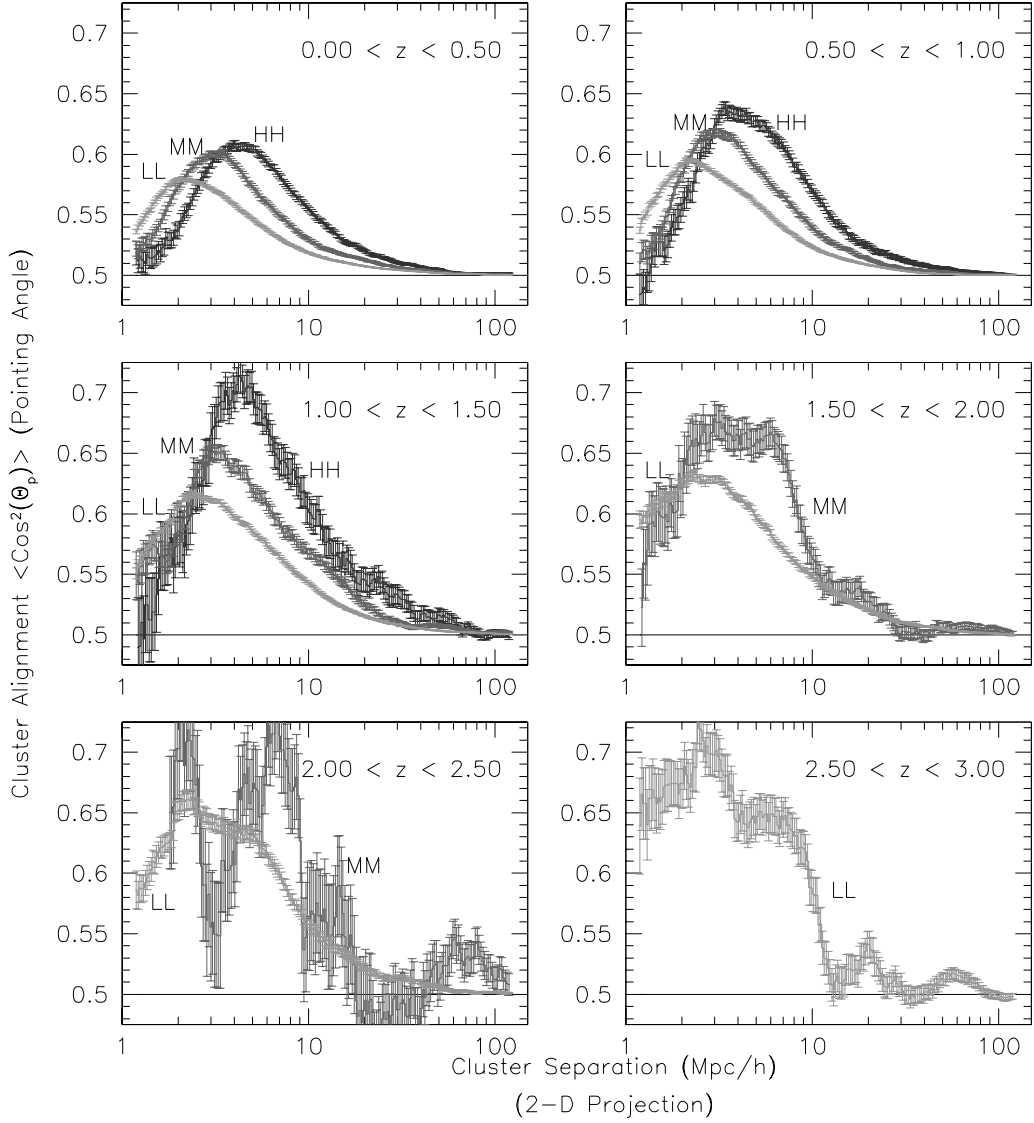}
    \caption{Same as Figure~\ref{cosvsrM}, but for the 
    2-D cluster alignment function (pointing angle method).
    \label{cosvsrM2d}}
\end{figure}
\begin{figure}
    \plotone{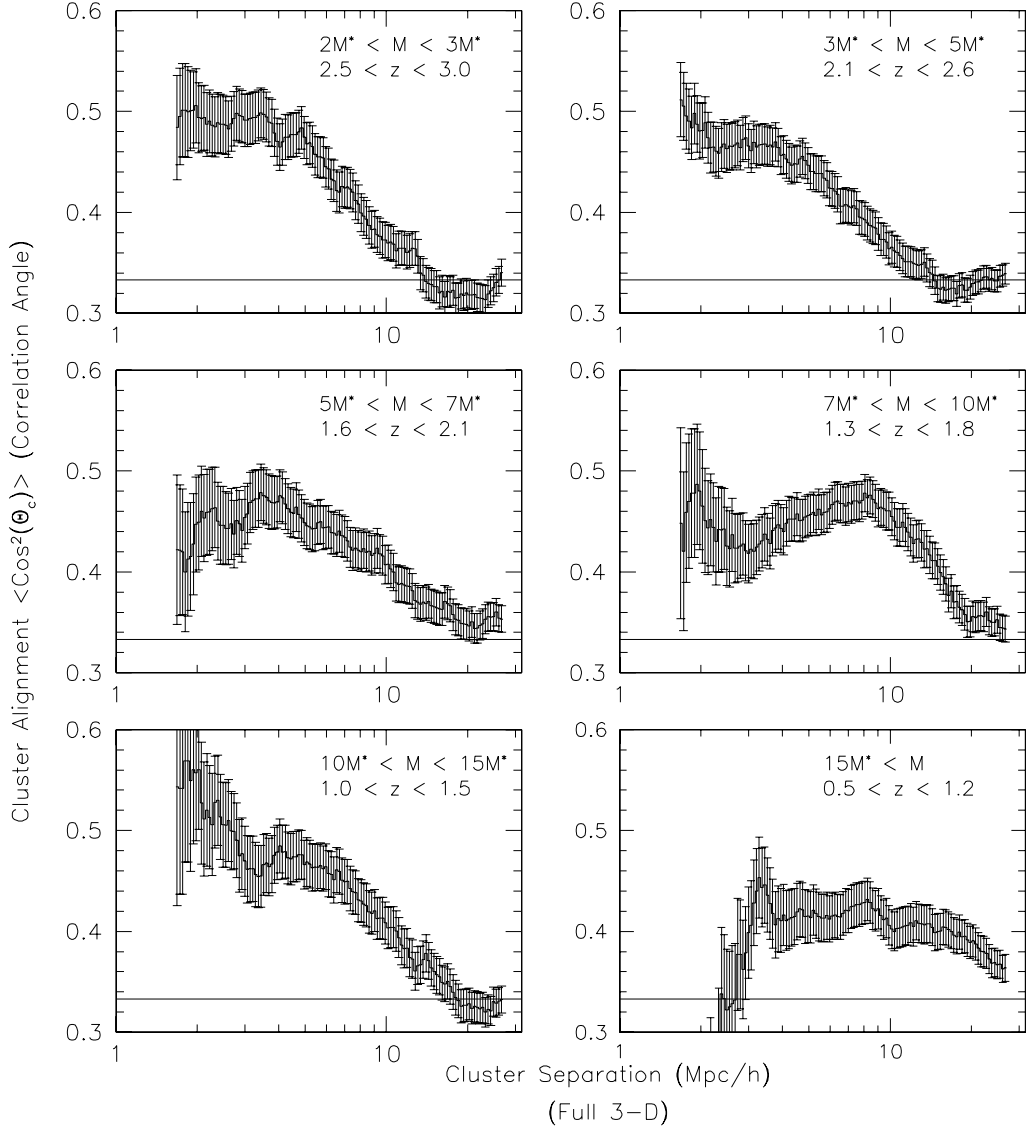}
    \caption{The 3-D cluster alignment for clusters of different masses shortly 
    after their formation. Clusters are divided into six mass bins, 
    and for each mass bin, cluster pairs are examined within the characteristic
    redshift range of their formation (see \S~\ref{sec:align}). 
    The mass and redshift ranges are shown
    in each panel (with ${\rm M}^{*}=10^{13}\,\mass$).
    \label{compareformation}}
\end{figure}
\begin{figure}
    \plotone{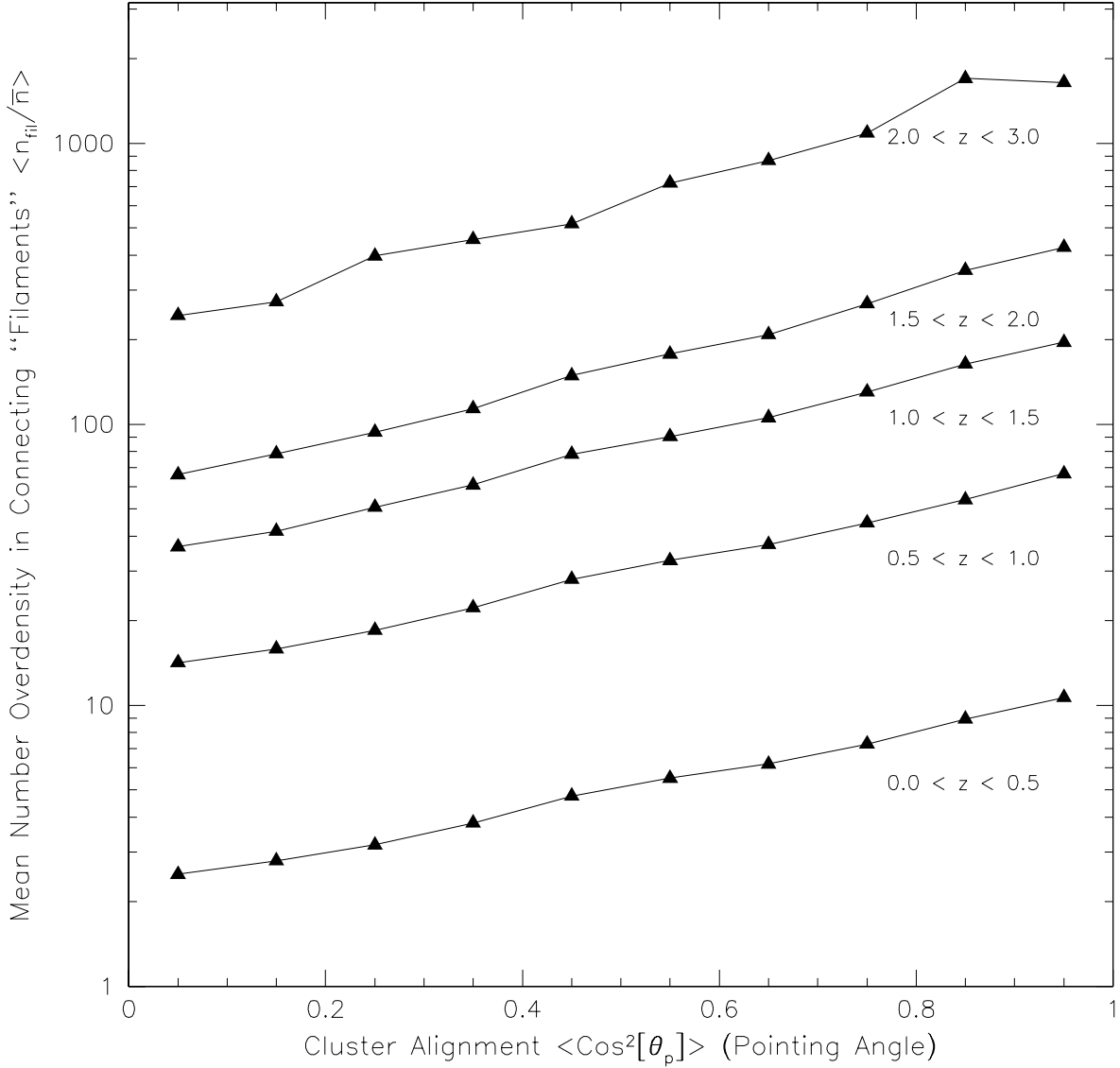}
    \caption{Mean number overdensity of clusters 
    within $1.5\,\dist$ of the line connecting
    cluster pairs is plotted as a function of pair alignment, for all pairs 
    with separations $r<50\,\dist$. The alignment, 
    $\ecosp$, is the average of the two pointing angles for
    a given pair of clusters, taken in bins of width 0.1. 
    Redshift increases from bottom to top, as labeled.
    \label{overdense}}
\end{figure}
\end{center}
\end{document}